# Change in the attenuation coefficient of light at the boundary between the troposphere and stratosphere

Kochin A. V.

Central Aerological Observatory,
Pervomaiskaya st. 3, Dolgoprudny, Moscow Region, Russia, 141701
Email: amarl@mail.ru

**Abstract**

The tropopause is the boundary between the troposphere and the stratosphere, so its height determines the nature of processes in the atmosphere and depends on them. The criterion for determining the height of the tropopause is a decrease in the vertical temperature gradient in a layer above 5 km and at least 2 km thick. Consequently, the uncertainty in determining the height of the section between the troposphere and the stratosphere is 2 km or more. In addition, the temperature profile changes due to the influence of many reasons, so sometimes the height of the tropopause cannot even be determined. The optical properties of air in the troposphere and stratosphere differ, which makes it possible to measure the height of the section with high accuracy using a radiosonde with an optical sensor. The launches of radiosondes with optical sensors made it possible to measure the height of the tropopause by changing the attenuation coefficient of visible light. It turned out that the lower edge of the tropopause corresponds to a sharp change in the attenuation coefficient. The repeatability of the results was confirmed by the simultaneous launch of radiosondes. The results of the work will be useful for the tasks of weather forecasting and climate research.

1. Introduction

The tropopause is a transitional layer between the troposphere and stratosphere. It has a complex structure, since the distribution of temperature, humidity and wind in it is determined by many factors. The height of the tropopause depends on the latitude of the place, the season of the year and the nature of atmospheric circulation. As the lower edge of the tropopause, a level above 5 km is taken, where the vertical temperature gradient decreases to 0.2 degrees / 100 m or lower, and the thickness of this layer is at least 2 km. According to this definition, the real height of the boundary between the troposphere and stratosphere air is located inside a sufficiently extended layer of the order of 2 km. The tropopause is a intercepting layer, which is due to the nature of the temperature distribution with altitude. As a intercepting layer, the tropopause prevents the transfer of aerosols and water vapor. Atmospheric pollutions accumulate under the tropopause. Therefore, visibility under the tropopause worsens, the sky takes on a whitish hue. The main attenuation occurs due to the absorption of light by aerosols and cloud particles in the troposphere. The source of aerosols is the earth's surface, so absorption depends on altitude. Ground measurements give only the value of integral attenuation and do not allow measuring the vertical profile of the attenuation coefficient. The optical properties of air in the troposphere and stratosphere differ, which makes it possible to measure the height of the section with high accuracy by changing the attenuation coefficient of light. The objective of the study was to establish a correspondence between the existing methods of calculating the height of the tropopause by the temperature profile and height to a sharp change in the attenuation coefficient.

## 2. Optical radiosonde

The study of optical processes in the atmosphere is carried out by actinometric radiosondes (Kondratiev and all, 1964; Kostyanoy, 1975; Asano and all, 2004; Nicoll & Harrison 2012, Philipona et all, 2012) with high-precision radiation receivers in various ranges. To ensure high-quality measurements, the radiosondes were equipped with a system for stabilizing the angle of inclination of the optical sensor of the radiosonde. The use of a stabilization system leads to a significant increase in the cost of the radiosonde. During the flight, the radiosonde is removed from the launch site at a distance of 100 km or more. Finding a landed radiosonde is a complex and expensive procedure, so a standard radiosonde is a disposable device. The installation of expensive sensors and stabilization systems on the radiosonde leads to an increase in the cost of sounding. This is a noticeable increase in costs, since one station produces about 800 launches of radiosondes per year for two-time sounding. To reduce the cost of the design, the optical sensor was implemented on a conventional commercial visible-range photodiode with a viewing angle of 60 degrees. Photodiodes were tested in a thermal chamber at a temperature of -70°C (Kochin 2021). The sensors were selected from a batch where the output signal did not depend on temperature. External view of a radiosonde is shown on Fig.1.

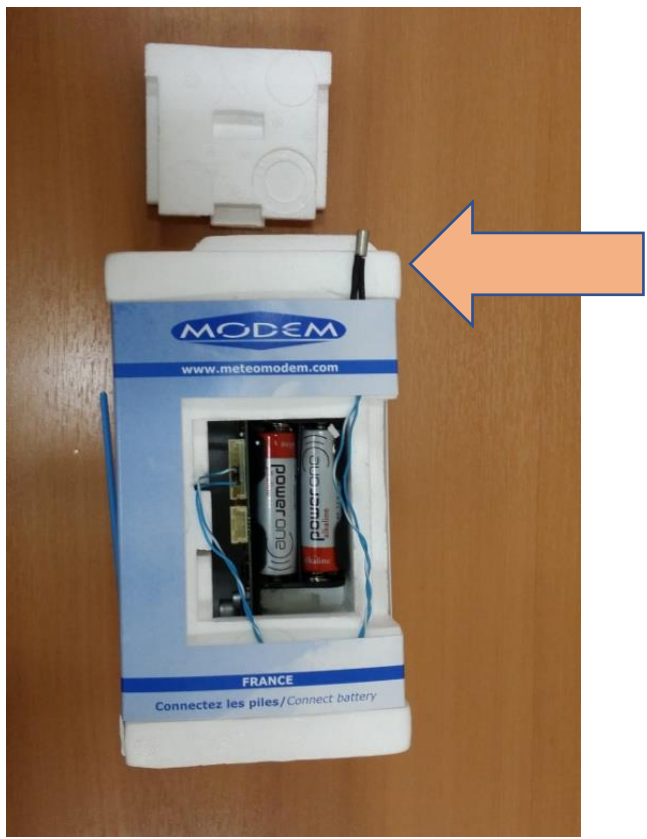

Fig.1. External view of a radiosonde with a visible light sensor. The sensor is marked with an arrow.

Experiments with optical sensors were carried out using M2K2DC radiosondes of the French company MODEM, which have an internal ADC with an analog signal input. The telemetry frequency was 1 Hz. The optical sensors were mounted in such a way that the maximum of the diagram was oriented along the line connecting the radiosonde and the balloon. The system oscillates like a pendulum during flight. The average angle of deviation from the vertical is 18 degrees. The average angle value was determined based on the results of experiments with a specialized radiosonde with a three-axis acceleration sensor (Dubovetsky 2015).

3. Observation results

Radiosondes with optical sensors in the visible range were launched at the standard time of 12 GMT+00 at the Dolgoprudnaya upper-air sounding station located in the Moscow region (55.93, 37.52). According to local time, the launch time was 15:00. The height of the Sun above the horizon varied from 30 to 50 degrees. More than 20 launches were carried out in various conditions (Kochin, 2021). The raw data is shown in Fig. 2.

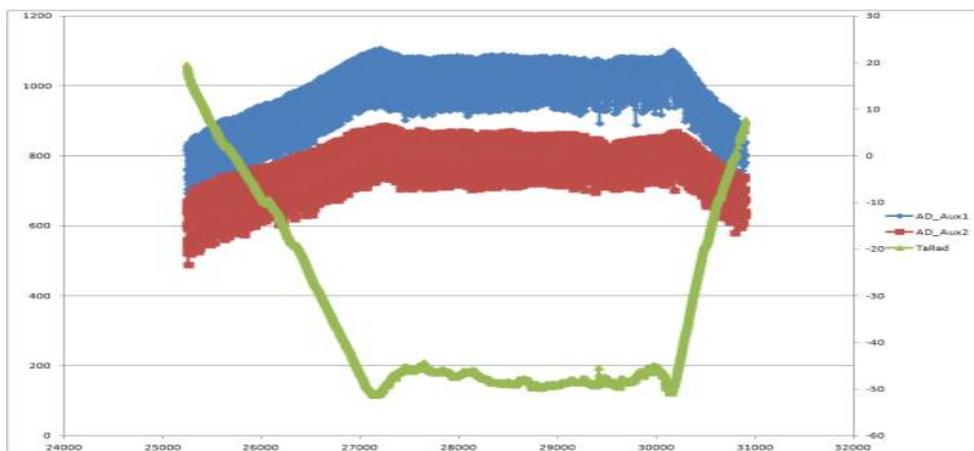

Fig.2. Data from the launch of a radiosonde with two sensors installed on it. The raw signal of the optical sensor of the visible range is shown in blue. The raw signal of the optical sensor in the near infrared range of the order of 1 microns is shown in red. Both sensors register the incident solar radiation. The green line shows the temperature. Height above sea level in meters on the horizontal axis. The left scale is the magnitude of the optical sensor signal in the ADC code. The right scale is the temperature, degrees Celsius.

Figure 2 shows the data of a special launch of a radiosonde with two sensors to compare the nature of signal changes in the visible and near-infrared ranges. The intensity of solar radiation is greater in the visible range, but the nature of the signal change does not depend on the wavelength. To confirm the coincidence of the results, double launches were carried out. The results are shown in Fig.3.

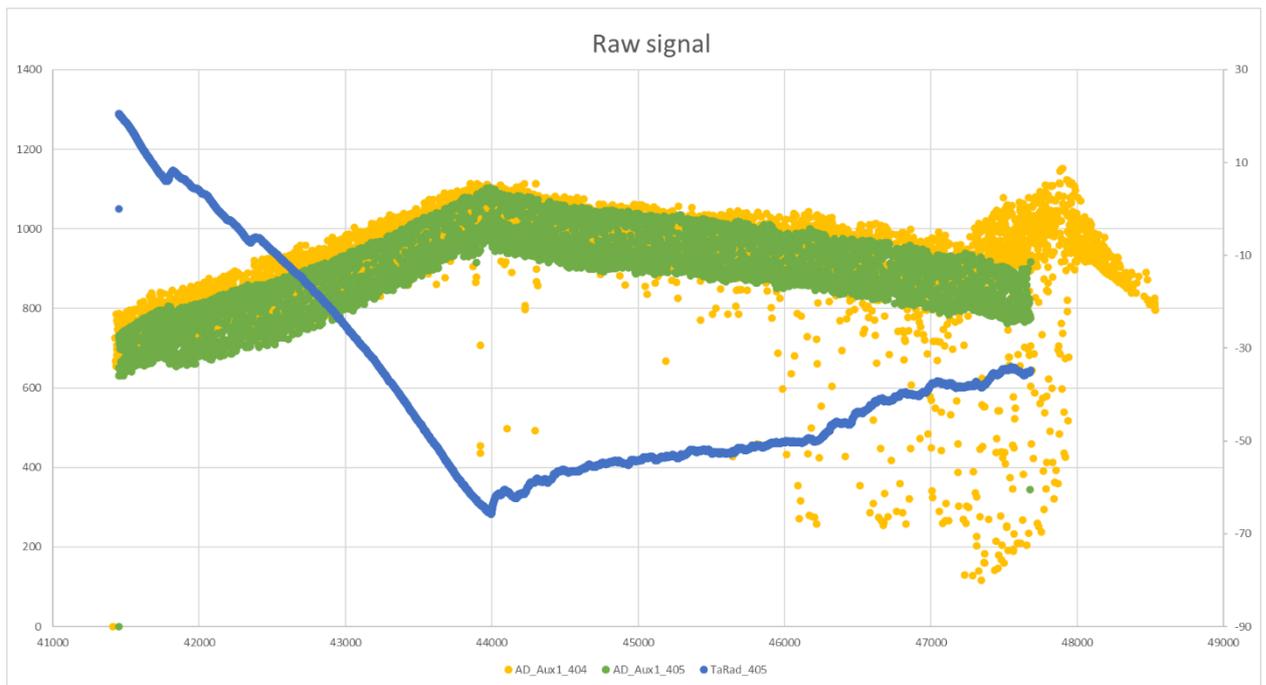

Fig. 3. Data of simultaneous launch of two radiosondes. Raw signals of optical sensors of radiosondes are shown in green and yellow. Both sensors register the incident solar radiation of the visible range. The blue line shows the temperature. Height above sea level in meters on the horizontal axis. The left scale is the magnitude of the optical sensor signal in the ADC code. The right scale is the air temperature, degrees Celsius.

In all launches, the nature of the change in the signals of the optical sensors of the radiosondes was the same. The signal increased to a certain height, and then decreased, as shown in Fig.2 and Fig.3. The increase in the signal is caused by a decrease in the optical thickness, which reduces the attenuation of solar radiation. Then the signal decreases due to a decrease in the height of the Sun, since the launches were made at 15 o'clock local time. The bend in the optical sensor signal always corresponded to the bend in temperature. In accordance with the criteria for choosing the tropopause, the bend corresponds to the height of the lower edge of the tropopause.

### 4. Profile of the attenuation coefficient of visible light in the atmosphere

The initial purpose of launching radiosondes with optical sensors was to measure the cloud top height (Kochin 2016). It was assumed that the use of simple sensors without a position stabilization system could not claim to correctly measure the vertical profile of the attenuation coefficient of visible light in the atmosphere. However, the nature of the change in the raw signal in Fig.2, the average level of which decreases from the tropopause to the Earth's surface by 30% (which is close to the known amount of attenuation of visible light in the atmosphere, taking into account the optical mass of the atmosphere at 15 o'clock local time), indicates some possibility of obtaining significant information.

In practice, the derivative of the average signal gave remarkable results. Before differentiation, the raw signal was adjusted to the height of the Sun, since the sounding began at 15 o'clock local time. After correction, the average signal level above the tropopause does not change. The raw signal fluctuates very much. Simple averaging does not give a satisfactory result, since the profile details are lost. After selecting the options, an empirical method for selecting signal maxima was chosen. In a sliding window of 30 values, the maximum signal was selected, and this value was written to a file. The resulting file was differentiated by height. The results are shown in Figure 4.

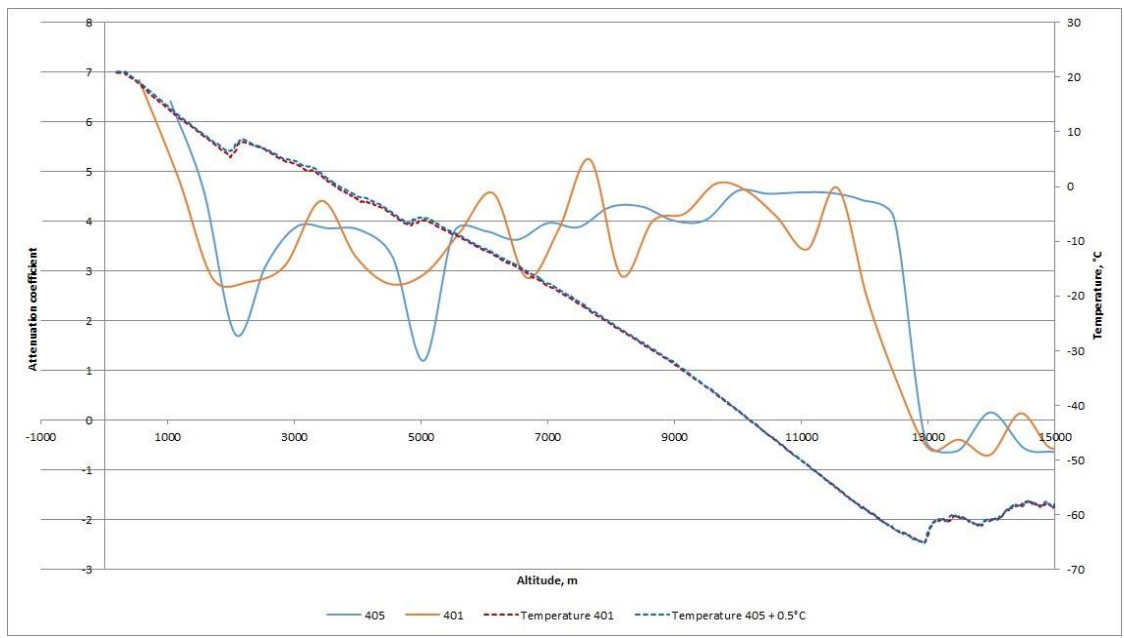

Fig. 4. Results of simultaneous launch of two radiosondes. Yellow shows the data of the radiosonde with a frequency of 401 MHz, blue shows the data of the radiosonde with a frequency of 405 MHz. The solid lines are the calculated attenuation coefficients, and the dotted lines are the temperature. The horizontal axes are height, the vertical scale on the left is the attenuation coefficient of $10^{-2}\text{km}^{-1}$, and the right scale is temperature, degrees Celsius.

Launches of radiosondes with optical sensors have shown the possibility of measuring the vertical attenuation profile. The obtained profiles have a peculiarity in the tropopause region, where the attenuation coefficient changes dramatically. To confirm the reliability of the obtained result, two radiosondes with optical sensors were launched simultaneously (Fig.2). The starting points were located at a distance of 1.5 km . At an altitude of about 13 km, corresponding to the height of the tropopause, both radiosondes synchronously registered a sharp decrease in the attenuation coefficient to almost zero (Fig.4). The difference in the height of the bend is almost not noticeable. The presence of a sharp change in the attenuation coefficient at the tropopause level has not been previously established. This change coincides with the position of the tropospheric inversion layer (Birner 2006).

Thus, the optical properties of the air below the tropopause differ sharply from the air above the tropopause. At an altitude of 2.5 and 5 km, sharp peaks are visible, which coincided with the inversion layers. This corresponds to the well-known ideas about a decrease in the concentration of aerosols in the inversion layer and an increase in their concentration below it, which causes changes in the transparency of atmospheric air. In the stratosphere, visible light practically does not weaken (Kochin 2021), and the main absorption occurs in the troposphere. The ratio between the height of the tropopause by the temperature profile and by the attenuation coefficient is shown in Fig. 5.

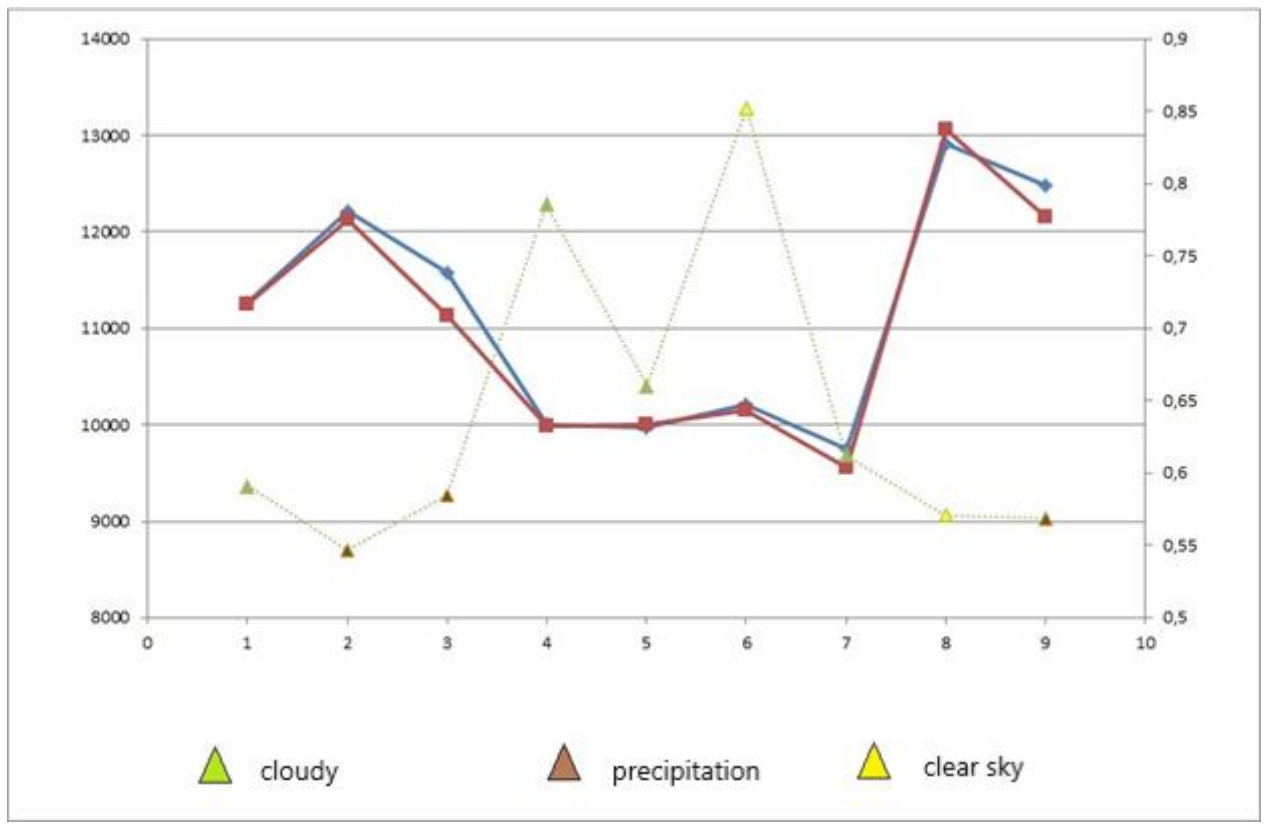

Fig. 5. The height of the tropopause according to the temperature profile and the attenuation coefficient. The red line shows the height of the tropopause by the temperature profile, the blue line shows the height of the tropopause by the attenuation coefficient. The dotted line shows the ratio of optical sensor signals at the surface and at an altitude of 20 km. Triangles show cases of clear skies (yellow), clouds without precipitation (green) and clouds with precipitation (brown).

To determine the ratio between the height of the tropopause by the temperature profile and by the attenuation coefficient, the results of 9 launches were processed (Fig.5). It turned out that these heights practically do not differ. Sometimes the height of the tropopause by the attenuation coefficient is higher by about 0.5 km. Thus, the existing method of calculating the height of the tropopause by the temperature profile correctly determines the distribution of aerosol concentration in the atmosphere, which is confirmed by experimental data.

Fig. 5 also shows the ratio of the signal of the optical sensor of the radiosonde on the surface to the signal at an altitude of 20 km. An increase in the height of the tropopause leads to an increase in attenuation in the troposphere, which can be seen by a decrease in this ratio. Clouds and precipitation change the magnitude of this ratio, but the height of the tropopause is the dominant cause. The fact of the presence of clouds is visible by the dispersion of the signal. In clouds with precipitation, the dispersion drops to zero (Kochin 2021). This result cannot yet be considered in quantitative terms, since the sensors have not been calibrated in power. The ratio of ADC codes may not correspond to the ratio of intensities. However, the qualitative type of dependence can serve as arguments for research in this direction.

6. **Conclusions**

The results of the work confirmed the ability of a radiosonde with an optical sensor to measure the height of the boundary between the air masses of the troposphere and stratosphere. This makes it possible to alternatively measure the height of the tropopause when it cannot be determined by temperature. Ease of use and low cost of consumables are the main factors for devices used on the network. The simplest sensors are used in this work. This will not significantly change the cost of sounding. The results of the work can be useful for weather forecasting and climate research.

**Acknowledgments.**

The author thanks the staff of the Central Aerological Observatory for their help in the work, and the staff of the Dolgoprudnaya upper-air station for launching radiosondes.